\documentclass[
    ,final            
  ]
  {aipproc}

\layoutstyle{6x9}

\begin{document}

\title{Wobbling motion in triaxial superdeformed nuclei}


\author{Masayuki Matsuzaki}{
  address={Department of Physics, Fukuoka University of Education,
Munakata, Fukuoka 811-4192, Japan}
}

\author{Yoshifumi R. Shimizu}{
  address={Department of Physics, Graduate School of Sciences,
Kyushu University, Fukuoka 812-8581, Japan}
}

\author{Kenichi Matsuyanagi}{
  address={Department of Physics, Graduate School of Science,
Kyoto University, Kyoto 606-8502, Japan}
}

\begin{abstract}
We discuss some characteristic features of the wobbling motion excited on the 
triaxial superdeformed Lu nucleus. We show how these features are connected to 
the moments of inertia microscopically calculated by means of the quasiparticle 
RPA in the rotating frame. 
\end{abstract}

\maketitle

 The concept of independent single particle motions in mean field potentials 
serves as a starting point of nuclear structure theories. Single particle 
energy levels vary as functions of various parameters of the mean field 
potential; bunch, degenerate, and form shell gaps. This is called shell 
structure. Shell structure is the clue to various nuclear phenomena. 
One of the most famous examples is the occurrence of the superdeformation 
with the axis ratio $2:1$ observed in the $A\sim150$ mass region. 
In this case we consider single particle energy levels in an axially 
symmerically deformed potential, for example the anisotropic harmonic 
oscillator one. These levels strongly degenerate at some deformations that 
correspond to simple integer axis ratios. Consequently shell gaps are 
formed. This occurs also in more realistic potentials. Since it is difficult 
for nucleons to excite across the gap, shell-closed configurations become 
stable. 

 A similar situation can be thought of in triaxially ($Y_{22}$) deformed 
potential when varying the triaxial parameter $\gamma$ from $0^\circ$ to 
$60^\circ$. Actually, a realistic energy surface calculation~\cite{ragnar} 
predicts the existence of stable configurations with $\epsilon_2\sim0.4$ and 
$\gamma\sim20^\circ$, where $\epsilon_2$ stands for a parametrization of the 
$Y_{20}$ deformation. When rotation sets in, the triaxial parameter requires 
three times larger range ($-120^\circ\le\gamma\le+60^\circ$ in the so-called 
Lund convention) according to the relation between the axis of rotation and 
that of deformation. Therefore $\gamma=+20^\circ$ and $-20^\circ$ in 
rotating systems represent different physical situations. Actually, according 
to Ref.~\cite{ragnar}, the energy minimum at $\gamma\simeq+20^\circ$ is 
stabler than that at $\gamma\simeq-20^\circ$. 

 The signal of triaxial deformation has long been sought for but the result 
has been ambiguous. From a theoretical viewpoint, however, Bohr and Mottelson 
predicted the existence of the nuclear wobbling motion in rapidly rotating 
triaxially deformed systems~\cite{bm}. This is a quantum analog of 
the one that has been known in classical mechanics~\cite{landau}.
A candidate of the configuration with $\epsilon_2\sim0.4$ and 
$|\gamma|\sim20^\circ$, which is called the triaxial superdeformation (TSD), 
has been known in a Lu isotope for years~\cite{lu0}. In 2001 an excited 
TSD band was reported for the first time in $^{163}$Lu~\cite{lu1}. 
In this work, extremely strong interband electric quadrupole transitions with 
about a hundred Weisskopf units were measured and therefore this was thought 
of as a clear evidence of a collective wobbling motion and consequently of 
triaxial deformation. 

\begin{figure}[htbp]
  \includegraphics[width=8cm]{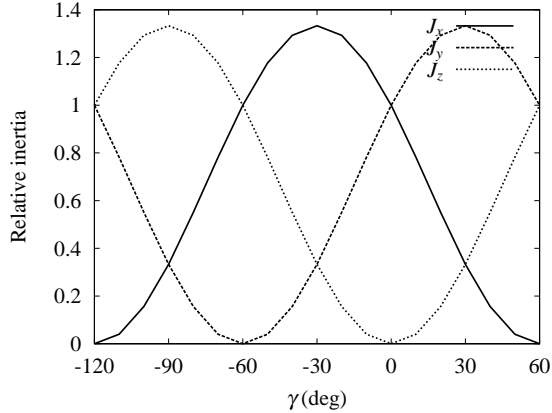}
  \caption{Irrotational model moments of inertia.}
  \label{fig1}
\end{figure}

 A characteristic feature of the wobbling motion is its excitation energy 
given by
\begin{equation}
\hbar\omega_\mathrm{wob}=\hbar\omega_\mathrm{rot}
\sqrt{\frac{\left(\mathcal{J}_x-\mathcal{J}_y\right)
   \left(\mathcal{J}_x-\mathcal{J}_z\right)}
     {\mathcal{J}_y
      \mathcal{J}_z}} \ ,
\label{dispbm}
\end{equation}
when we name the axis of the main rotation the $x$ axis. Here $\omega_\mathrm{rot}$ 
is the rotational frequency of the main rotation and $\mathcal{J}$s are 
moments of inertia. In order for $\omega_\mathrm{wob}$ to be real, the order 
of $\mathcal{J}$s is constrained; for example 
$\mathcal{J}_x>\mathcal{J}_y,\mathcal{J}_z$ or
$\mathcal{J}_x<\mathcal{J}_y,\mathcal{J}_z$. In the case of the rotor model, 
$\mathcal{J}$s are given by hand. As the input, a few models of nuclear 
moments of inertia are known. Among them, aside from its magnitude, the 
$\gamma$ dependence in the irrotational model,
\begin{equation}
\mathcal{J}_k^\mathrm{irr}=4B\beta^2\sin^2{(\gamma+\frac{2}{3}\pi k)} ,
\label{irr}
\end{equation}
with $k =$ 1 -- 3 denoting the $x$ -- $z$ principal axes, $B$ the irrotational 
mass parameter, and $\beta$ a deformation parameter similar to $\epsilon_2$, 
is believed to be appropriate for the collective motion. Note that the overall 
magnitude is not relevant to Eq.~(\ref{dispbm}). Its $\gamma$ dependence is 
shown in Fig.~\ref{fig1}. This figure suggests choosing $\gamma\simeq-20^\circ$ 
out of $|\gamma|\simeq20^\circ$ because of 
$\mathcal{J}_x>\mathcal{J}_y>\mathcal{J}_z$, whereas the potential energy 
surface calculation mentioned above suggests $\gamma\simeq+20^\circ$. 
In order to solve this puzzle, we need a framework that determines 
$\mathcal{J}$s microscopically. 

 We adopt the cranking model plus random phase approximation (RPA) in the 
manner of Marshalek~\cite{ma}. The one-body Hamiltonian is given by 
\begin{equation}
h'=h-\hbar\omega_\mathrm{rot}J_x \ ,
\end{equation}
here $h$ denotes the Nilsson plus BCS Hamiltonian. We perform the RPA to the 
residual pairing plus quadrupole interaction. According to signature 
($\pi$ rotation about the $x$ axis) symmetry, only the $\alpha=1$ 
($r=\exp{[-i\pi\alpha]}=-1$) sector of the interaction, 
\begin{equation}
H_\mathrm{int}^{(-)}=-\frac{1}{2}\sum_{K=1,2} \kappa_K^{(-)} Q_K^{(-)\dagger} Q_K^{(-)} \ ,
\end{equation}
is relevant for the description of the wobbling motion. The equation of motion, 
\begin{equation}
\left[h'+H_\mathrm{int}^{(-)},X_n^\dagger\right]_\mathrm{RPA}
=\hbar\omega_n X_n^\dagger \ ,
\end{equation}
for the $n$-th eigenmode $X_n^\dagger$ leads to a pair of coupled equations for the 
transition amplitudes. Then, by assuming $\gamma\neq0$, this can be cast into the 
form~\cite{ma}, 
\begin{equation}
\hbar\omega_\mathrm{wob}=\hbar\omega_\mathrm{rot}
\sqrt{\frac{\left(\mathcal{J}_x-\mathcal{J}_y^\mathrm{(eff)}(\omega_\mathrm{wob})\right)
   \left(\mathcal{J}_x-\mathcal{J}_z^\mathrm{(eff)}(\omega_\mathrm{wob})\right)}
     {\mathcal{J}_y^\mathrm{(eff)}(\omega_\mathrm{wob})
      \mathcal{J}_z^\mathrm{(eff)}(\omega_\mathrm{wob})}} \ ,
\label{disp}
\end{equation}
for $n=\mathrm{wob}$. See Ref.~\cite{msmf} for details. 
The form of Eq.~(\ref{disp}) is evidently parallel to Eq.~(\ref{dispbm}) but here 
$\mathcal{J}_{y,z}^\mathrm{(eff)}(\omega_\mathrm{wob})$
are dynamical ones that are determined simultaneously with $\omega_\mathrm{wob}$. 
In this sense Eq.~(\ref{disp}) is a highly nonlinear equation and it is not 
trivial whether a collective wobbling solution is obtained from it or not. 

\begin{figure}[htbp]
  \includegraphics[width=7.5cm]{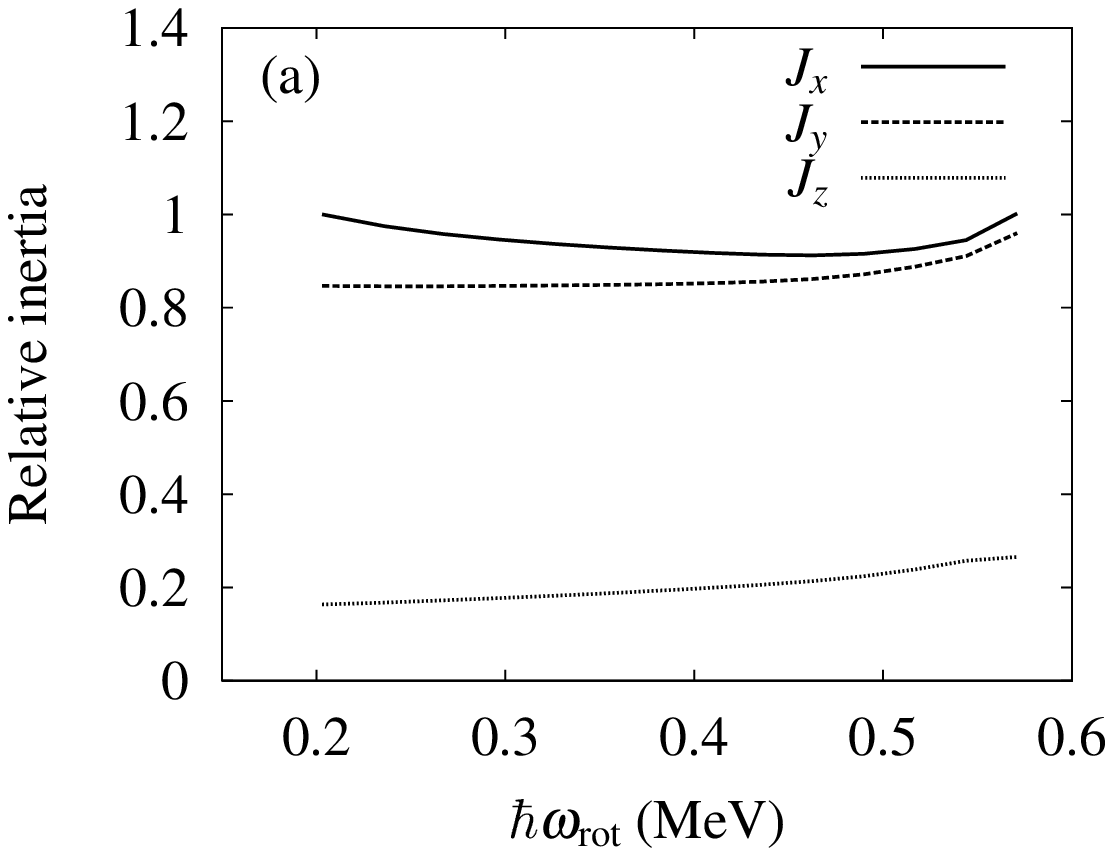}
  \includegraphics[width=7.5cm]{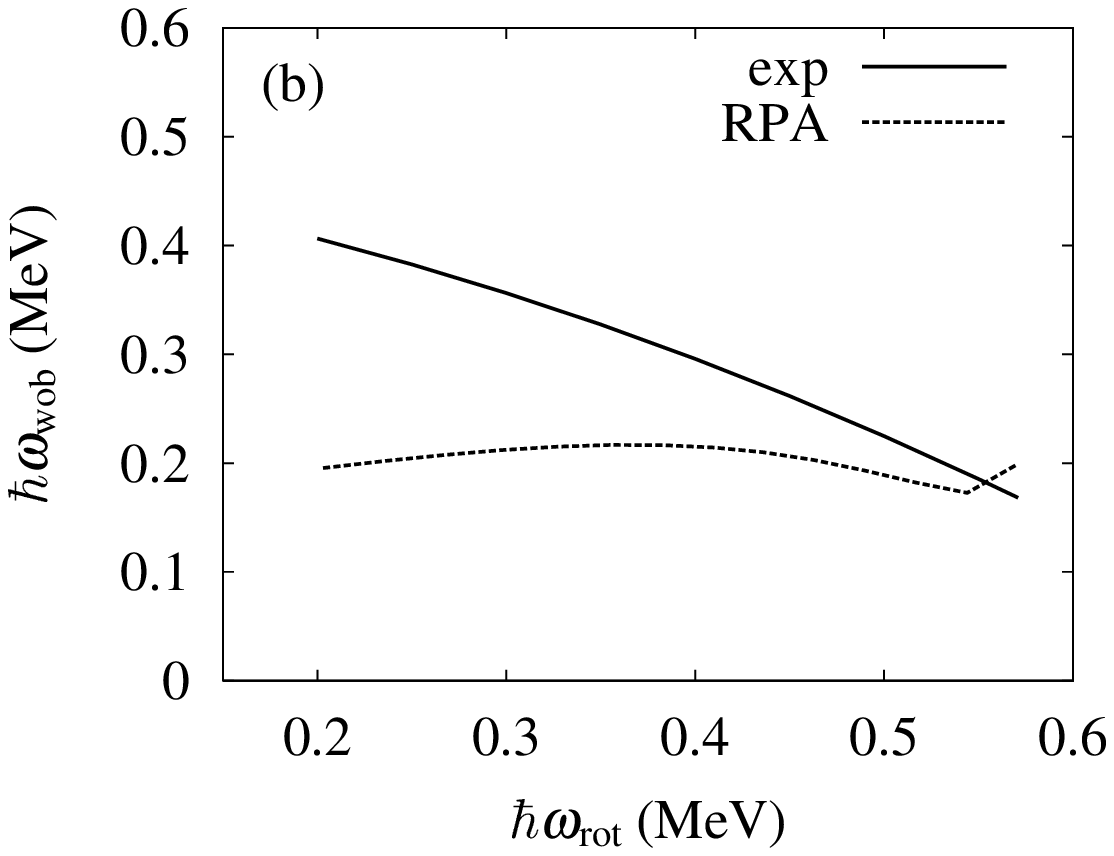}
\caption{(a) Calculated moments of inertia, and (b) experimental and calculated 
excitation energies of the wobbling mode in $^{163}$Lu. 
Note that the proton $BC$ crossing occurs at around 
$\hbar\omega_\mathrm{rot}\geq$ 0.55 MeV in the calculation. Data are taken from 
Refs.~\cite{lu1,lu2}. (Taken from Ref.\cite{msmr}.)}
  \label{fig2}
\end{figure}

\begin{figure}[htbp]
  \includegraphics[width=8cm]{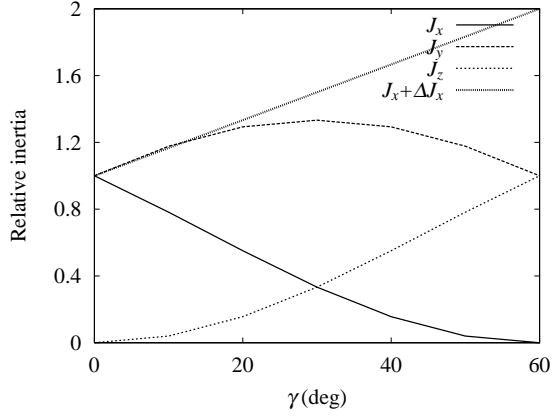}
  \caption{Schematic drawing of the alignment contribution from the last odd 
quasiparticle to the moment of inertia.}
  \label{fig3}
\end{figure}

 We obtained an extremely collective solution for $^{163}$Lu by adopting 
$\epsilon_2=0.43$, $\gamma=+20^\circ$, and $\Delta_n=\Delta_p=0.3$ MeV. 
The result is shown in Fig.~\ref{fig2}. Figure \ref{fig2}(a) graphs the moments 
of inertia. This figure indicates $\mathcal{J}_x>\mathcal{J}_y>\mathcal{J}_z$. 
Then what is the relation to the irrotational $\gamma$ dependence 
that is believed to be realistic? 
The key is that $^{163}$Lu can be regarded as the system consisting 
of the collective rotor and one quasiparticle. Contrastively to the case of the 
particle rotor model in which only the rotor is responsible for the moment of 
inertia, in the present model the last odd quasiparticle also bears inertia. 
Thus, the calculated moments of inertia in Fig.~\ref{fig2}(a) can be interpreted 
as a superposition of an irrotational-like one ($\mathcal{J}_x<\mathcal{J}_y$) 
of the rotor and an additional alignment contribution (mainly to $\mathcal{J}_x$) 
from the last odd quasiparticle as schematically depicted in Fig.~\ref{fig3}.

 Next we discuss the $\omega_\mathrm{rot}$ dependence of $\omega_\mathrm{wob}$ 
presented in Fig.~\ref{fig2}(b). When the moments of inertia are independent 
of $\omega_\mathrm{rot}$, $\omega_\mathrm{wob}$ is proportional to 
$\omega_\mathrm{rot}$. This in turn indicates that the actual moments of inertia 
depend on $\omega_\mathrm{rot}$. Figure \ref{fig2}(a) shows that the calculated 
$\mathcal{J}$s do depend on $\omega_\mathrm{rot}$. Seemingly their dependence 
is weak, the decrease of $\mathcal{J}_x-\mathcal{J}_y^\mathrm{(eff)}$ makes 
$\omega_\mathrm{wob}$ a flat or decreasing function. 

\begin{figure}[htbp]
  \includegraphics[width=8cm]{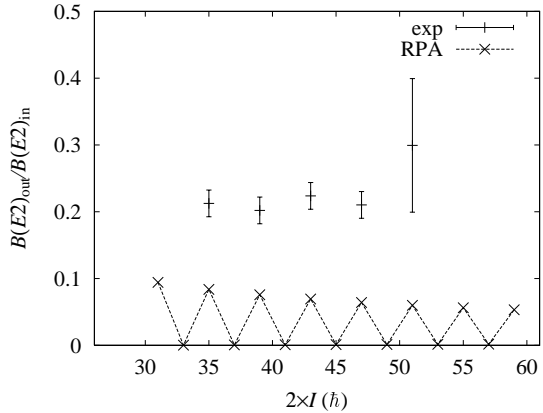}
 \caption{Interband $E2$ transition rates for 
$I$ (wobbling) $\rightarrow$ $I\pm1$ (yrast) transitions as functions of 
$2\times$ spin $I$. They are divided by 
the in-band $E2(I \rightarrow I-2)$ transition rates. Experimental values 
were taken from Ref.~\cite{lu2}. 
Noting that the states $I+1$ (yrast) are slightly higher in energy than 
$I$ (wobbling) at $I>51/2$ and 
$B(T_\lambda;I\rightarrow I+1)\simeq B(T_\lambda;I+1\rightarrow I)$ 
at high spins, we plotted those for $I\rightarrow I+1$ at the places with the 
abscissae $I+1$ in order to show clearly their characteristic staggering 
behavior. (Taken from Ref.\cite{msmr}.)}
  \label{fig4}
\end{figure}

\begin{figure}[htbp]
  \includegraphics[width=6.5cm]{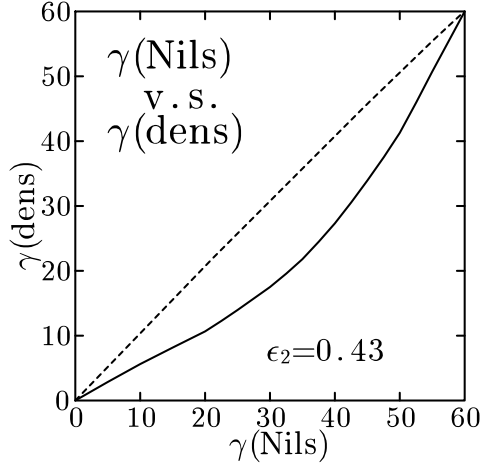}
\caption{Relation between $\gamma$ of the Nilsson potential and that of the 
density distribution of the nucleus (solid curve).}
  \label{fig5}
\end{figure}

 Now we come to the $B(E2)_\mathrm{out}/B(E2)_\mathrm{in}$ ratio. Here 
$B(E2)_\mathrm{out}$ means the reduced transition rate of the interband 
electric quadrupole transition between the wobbling and the yrast TSD bands, 
while $B(E2)_\mathrm{in}$ means that of the in-band one. Therefore this ratio 
measures the collectivity of the wobbling excitation. Figure \ref{fig4} 
compares the experimental and the theoretical ratios. Although in this 
calculation the calculated ones are factor 2 -- 3 smaller, recently we found 
its reason. The origin of the discrepancy is the difference of the physical 
meaning of the triaxial parameter $\gamma$. We adopted in the above 
calculation $\gamma=+20^\circ$ of the Nilsson potential but we found that the 
discrepancy is resolved if $\gamma=+20^\circ$ of the density distribution is 
adopted because their relation is not a diagonal straight line 
(Fig.~\ref{fig5}). 
That is, $\gamma(\mathrm{dens})=+20^\circ$ corresponds to 
$\gamma(\mathrm{Nils})\simeq+30^\circ$, and larger $\gamma(\mathrm{Nils})$ 
leads to larger $B(E2)_\mathrm{out}$. 
This will be discussed in detail in Ref.~\cite{smfc}. 

\begin{figure}[htbp]
  \includegraphics[width=8cm]{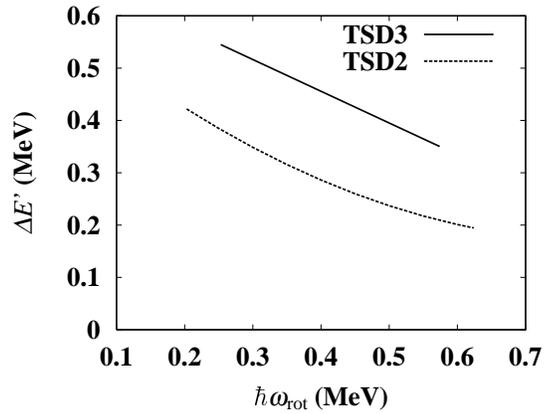}
  \caption{Experimental excitation energies of the two phonon (TSD3) and the one phonon 
(TSD2) wobbling states relative to the yrast triaxial superdeformed (TSD1) states in 
$^{163}$Lu. Data are taken from Ref.~\cite{2phonon}. (Taken from Ref.~\cite{mo}.)} 
  \label{fig6}
\end{figure}

 Finally we mention the anharmonicity in the observed wobbling spectrum. 
In Ref.~\cite{2phonon} the two phonon wobbling excitation was reported. The 
data exhibits strong anharmonicity as shown in Fig.~\ref{fig6}. This might 
indicate softness of the potential energy surface. As a numerical experiment 
we examined a calculation for $^{162}$Yb, which is the system with the last 
odd quasiparticle in $^{163}$Lu removed. In this nucleus we did not obtain a 
wobbling solution. This result is quite natural because this nucleus does 
not have the last odd quasiparticle that produces the additional contribution 
to $\mathcal{J}_x$ and that consequently leads to the existence of the wobbling 
motion by making $\mathcal{J}_x-\mathcal{J}_y^\mathrm{(eff)}>0$. 
Actually we confirmed that the angular momentum vector is tilted in this nucleus 
following the instability of the wobbling motion~\cite{mo}. 

 To summarize, we have discussed some characteristics of the one phonon and the 
two phonon wobbling excitations in the triaxial superdeformed nucleus, $^{163}$Lu, 
which are 
determined by the behavior of the moments of inertia. First we have shown 
that the wobbling motion in positive $\gamma$ nuclei emerges thanks to the 
alignment contribution to the moment of inertia superimposed on the collective 
contribution. Second we have discussed that the decreasing behavior of the 
observed excitation energy of the one phonon wobbling is brought about by 
the rotational frequency dependence of the dynamical moments of inertia. 
Possible $\omega_\mathrm{rot}$ dependence of $\gamma$ would also affect 
$\omega_\mathrm{wob}$. Thirdly we have pointed out the importance of 
self-consistency between density and potential in determining an appropriate value 
of $\gamma$ that determines the transition strength. 
Finally we have discussed a possible ``phase transition" to the tilted axis 
rotation regime, associated with the instability with respect to 
the wobbling degree of freedom.

\end{document}